\documentclass[conference]{IEEEtran}
\IEEEoverridecommandlockouts
\usepackage{cite}
\usepackage{url}  
\usepackage{hyperref}

\usepackage{lscape}
\usepackage{hologo}
\usepackage{amsmath,amssymb,amsfonts}
\usepackage{algorithmic}
\usepackage{graphicx}
\usepackage{textcomp}
\usepackage{xcolor}
\usepackage{flushend}
\usepackage{float}
\usepackage{diagbox}
\usepackage{makecell}
\usepackage{booktabs} 
\def\BibTeX{{\rm B\kern-.05em{\sc i\kern-.025em b}\kern-.08em
    T\kern-.1667em\lower.7ex\hbox{E}\kern-.125emX}}
\begin{document}


\title{Electrodermal Insights into Stress Dynamics of AR-Assisted Safety Warnings 
in \\Virtual Roadway Work Zone Environments}



\author{
    \IEEEauthorblockN{Fatemeh Banani Ardecani}
    \IEEEauthorblockA{
        \textit{William States Lee College of Engineering} \\
        \textit{University of North Carolina at Charlotte} \\
        USA \\
        fbanania@charlotte.edu}
    \and
    \IEEEauthorblockN{Omidreza Shoghli\textsuperscript{*}\thanks{\textsuperscript{*}Corresponding Author}}
    \IEEEauthorblockA{
        \textit{William States Lee College of Engineering} \\
        \textit{University of North Carolina at Charlotte} \\
        USA \\
        oshoghli@charlotte.edu}
}
\maketitle

\begin{abstract}
This study examines stress levels in roadway workers utilizing AR-assisted multi-sensory warning systems under varying work intensities. A high-fidelity Virtual Reality environment was used to replicate real-world scenarios, allowing safe exploration of high-risk situations while focusing on the physiological impacts of work conditions. Wearable sensors were used to continuously and non-invasively collect physiological data, including electrodermal activity to monitor stress responses. Analysis of data from 18 participants revealed notable differences in EDR between light- and medium-intensity activities, reflecting variations in autonomic nervous system activity under stress. Also, a feature importance analysis revealed that peak and central tendency metrics of EDR were robust indicators of physiological responses, between light- and medium-intensity activities. The findings emphasize the relationship between AR-enabled warnings, work intensity, and worker stress, offering an approach to active stress monitoring and improved safety practices. By leveraging real-time physiological insights, this methodology has the potential to support better stress management and the development of more effective safety warning systems for roadway work zones. This research also provides valuable guidance for designing interventions to enhance worker safety, productivity, and well-being in high-risk settings.

\end{abstract}


\section{Introduction}

Roadway work zones, essential for transportation infrastructure maintenance and development, continue to pose significant safety risks despite ongoing improvements in safety measures and technologies. Recent data reveals a concerning trend in work zone fatalities. In 2021, work zone fatalities rose to 956, an 11\% increase from 2020, which recorded 863 deaths \cite{atssa2023workzone} and in 2022 alone an alarming increase in fatalities, with 2,066 transportation sector fatalities \cite{bls2023cfoi}. This ongoing trend highlights persistent challenges in mitigating hazards for workers within work zones.
Existing research on the physical demands of roadway work zones highlight specific factors that intensify stress levels among workers. These include lane closures, proximity to moving traffic, night shifts, and the operation of heavy equipment. Such conditions adversely affect cognitive performance, decision-making abilities, and overall safety \cite{sabeti2023mad, lim2017analyzing}. While qualitative studies have explored strategies to mitigate work stress \cite{van2000work}, stress among roadway construction workers not only elevates the likelihood of errors, injuries, and health complications but also negatively impacts productivity levels \cite{jebelli2018eeg}. These combined factors highlight the pressing need for innovative interventions to enhance worker safety and efficiency in high-stress construction environments. Augmented Reality (AR) technology shows significant potential for enhancing worker alerts, enabling faster and more effective responses to work zone hazards \cite{sabeti2024augmented}. 
\begin{figure*}[!htb]
    \centering
    \includegraphics[width=0.82\textwidth]{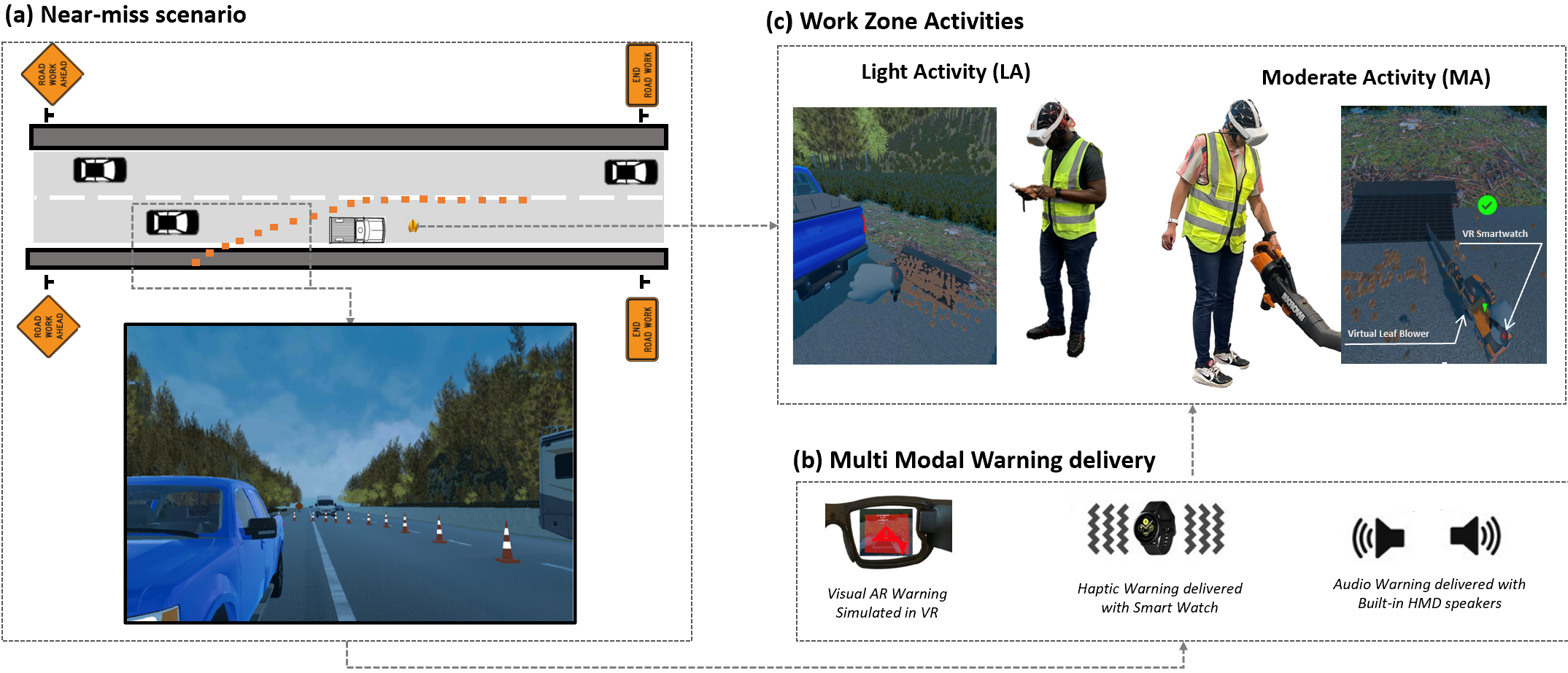}
    \caption{Overview of (a) Near Miss scenario in the simulated work zone layout, (b) components of the multi-modal warning delivery system, and (c) immersive VR environment for light and moderate activity.}
    \label{fig_1}
\end{figure*}
Stress impacts individuals differently, resulting in varied physiological responses \cite{ardecani2024assessing}. Electrodermal Activity (EDA), also known as Galvanic Skin Response (GSR), has been utilized as a physiological indicator of stress and arousal due to its sensitivity to the sympathetic branch of the autonomic nervous system \cite{benedek2010continuous, poh2010wearable}. EDA reflects the skin's electrical properties, influenced by sweat gland activity and controlled by the sympathetic nervous system. This exclusivity makes it a reliable stress indicator, unlike measures such as heart rate or skin temperature, which can also be affected by parasympathetic activity \cite{braithwaite2013guide, picard2016multiple}. Previous studies have highlighted EDA's role in monitoring stress across diverse settings, including emotional content reading \cite{sarinana2017cooperation}, underwater scenarios \cite{posada2018electrodermal}, and construction sites where it has been used to assess workers' stress levels and perceived risks in real time \cite{shakerian2021assessing, choi2019feasibility}. These findings underscore EDA’s utility in safety research, particularly as it remains unaffected by parasympathetic interference, offering unique insights into stress and risk perception in occupational and high-risk environments \cite{zou2017emotional}. EDA signals are typically decomposed into: (1) the tonic component (EDL), reflecting general arousal and long-term stress levels, and (2) the phasic component (EDR), capturing short-term responses to hazard-related stimuli \cite{greco2015cvxeda, greco2016force}. 
Highway worker fatalities highlight the critical need for innovative safety measures in roadway work zones. While Augmented Reality (AR) shows potential for enhancing worker alerts \cite{sabeti2022toward}, its impact on stress has not been fully explored. To address this gap, the objective of this study is to analyze stress responses in workers performing light- and medium-intensity tasks while receiving multi-sensory AR-enabled safety warnings. Building on existing literature, that identifies physical activity intensity \cite{snyder2022aperture} and cognitive load as key factors influencing worker stress \cite{ardecani2025neural}, the research utilizes a high-fidelity virtual reality environment to safely simulate realistic, high-risk scenarios. Additionally, by developing a non-invasive approach for continuous stress monitoring, this study aims to enhance safety practices, improve worker well-being, and increase productivity in roadway work zones.

\section{Method}
To fulfill the study's objective, an experiment was conducted in a high-fidelity Virtual Reality (VR) environment designed to replicate a real-world roadway work zone. The study protocol (21-0357) was approved by the Institutional Review Board (IRB) at the University of North Carolina at Charlotte. Participants carried out two routine roadway maintenance tasks in the VR environment while receiving simulated AR warnings, as shown in Figure \ref{fig_1}. We focused on understanding workers' stress responses after receiving AR-assisted warnings to evaluate the usability and effectiveness of AR-assisted safety technologies.

To achieve this, we evaluated EDA using a wristband-style wearable device equipped with an EDA sensor, enabling precise synchronization of physiological signals. EDA was measured through variations in skin conductance driven by sweat gland activity, reflecting sympathetic nervous system responses. This approach highlights the utility of EDA as an effective indicator of stress and arousal, suitable for both controlled experimental studies and real-world applications.

\subsection{Experiment Design and Procedure}

\textit{Experiment Design}: The experiment was designed to involve participants performing maintenance activities categorized into light and moderate intensity, as shown in Figure \ref{fig_1}(c). While participants performed each task, warnings were delivered at specific time intervals to assess their responses under varying conditions of physical and cognitive demand. These warnings, designed to communicate potential risks, were delivered simultaneously through haptic, audio, and visual cues. The haptic warnings were implemented using the Tizen Native framework \cite{tizenIoT}, which utilized predefined patterns available in its API. Audio warnings consisted of high-pitched beeps with a frequency of 44,100 Hz and a duration of 0.2 milliseconds. Visual warnings were presented through the functionality of augmented reality (AR) within the virtual reality (VR) environment.

Maintenance tasks were divided into low and moderate-intensity activities, as shown in Figure \ref{fig_1}(c), following approaches used in prior studies based on the Energy-Expenditure Prediction Program (EEEP) \cite{jebelli2019application, virokannas1996thermal}. For this study, light activity involved inspecting a clogged stormwater inlet by taking photographs of the blockage near the roadway shoulder. The moderate activity task in this study involved a common maintenance activity, requiring participants to clean the clogged stormwater inlet using a leaf blower in roadway work zones, where debris, vegetation, and leaves often obstruct curbs.

\textit{Experiment Procedure}: First, the purpose of the study and the roles of the participants were clearly communicated to the volunteers. After obtaining informed consent, each session was structured to be completed within a maximum duration of 30 minutes, and each task continued until all designated warnings were delivered and the administrator indicated that the session was completed.

\subsection{Apparatus}
\begin{figure}[!htb]
    \centering
    \includegraphics[width=0.45\textwidth]{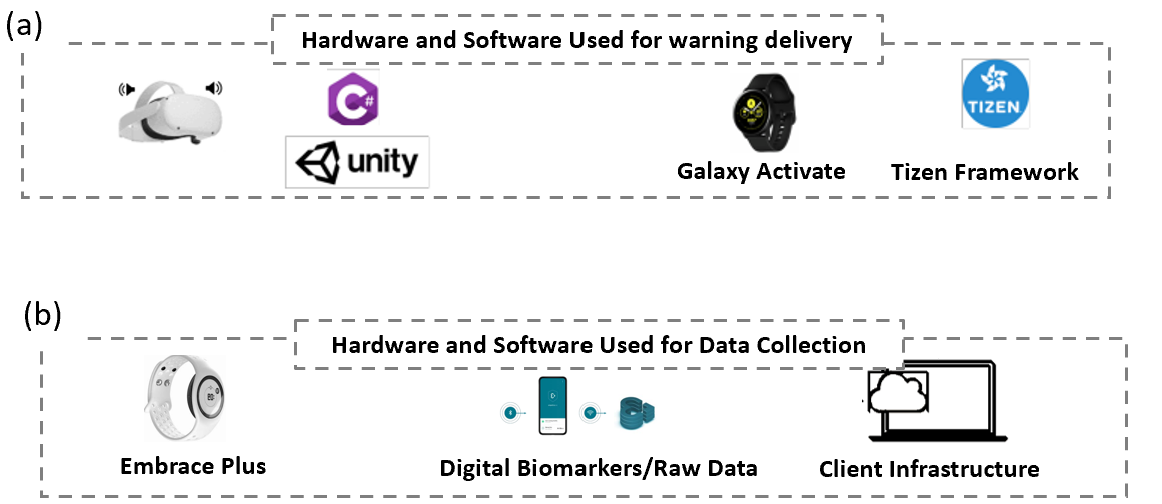}
    \caption{Hardware and software used for (a) warning delivery and (b) data collection.}
    \label{fig_2}
\end{figure}

\textit{Roadway Work Zone in Virtual Reality:}
The study utilized guidelines from the Manual on Uniform Traffic Control Devices \cite{texasMUTCD2006} to create a virtual highway work zone environment. The simulated setup, illustrated in Figure \ref{fig_1}, featured live traffic and detailed 3D models, closely replicating a real-world roadway work zone. Figure \ref{fig_1}(c) showcases the immersive VR environment from the participants' perspective as they carried out experiment tasks, developed by the research team \cite{sabeti2024augmented}. Participants were equipped with a VR headset and performing tasks using real-life tools:

\textit{Oculus Quest 2 VR headset and Warning Watch:} As shown in Figure \ref{fig_2}(a), Oculus Quest 2 VR headset was employed for virtual reality simulation. The Unity 3D game engine was utilized to develop the VR interface, ensuring an interactive and realistic user experience. Also, participants wore a Samsung Galaxy Watch, to receive haptic warnings delivered through the Tizen Native framework.

\textit{Leaf Blower and Tablet:} For task-specific activities, as shown in Figure \ref{fig_1}(c), participants used an Amazon Fire Tablet to capture photographs of a clogged stormwater inlet during the light activity. For the moderate activity, participants operated a WORX WG509 12Amp leaf blower to remove debris and clear the drainage area. 

\textit{Physiological Sensing Wristband:}
The Embrace Plus wristband \cite{embraceplus}, as shown in Figure \ref{fig_2}(b), an off-the-shelf wearable device, was utilized to measure participants' EDA signals. These signals were recorded concurrently, with EDA at a frequency of 4 Hz.

As illustrate in Figure \ref{fig_1}(b), participants could view and interact with the warning smartwatch, tablet, and leaf blower within the VR environment while simultaneously using them in the real-world setting during the experiment.

\subsection{Participants}

The study analyzed data from 18 participants. Among the final sample (N=18), the participants had an average age of 28.27 years. Five participants reported no prior work experience, while the remaining 13 participants had an average of 3.92 years of work experience (SD = 4.78). On average, participants completed the medium-intensity tasks in 1 minute and 42 seconds, while light-intensity tasks took 1 minute and 59 seconds.

\subsection{Physiological Data Analysis}

We developed a three-step data analysis framework to process raw Electrodermal Activity (EDA) data. This framework consists of data cleaning, to remove artifacts and noises; data analysis, to process and extract the key physiological features; and stochastic inference, to interpret the refined data and provide insights into underlying physiological responses. 

Data cleaning: Initially, a low-pass Butterworth filter 10 Hz was applied to attenuate the most common high-frequency noise, such as environmental artifacts, sensor motion artifacts, muscle movement artifacts \cite{guo2024unveiling}. Similarly, a second-order high-pass filter (Hamming window, cut-off frequency fc =0.05Hz) was employed to remove baseline drift and eliminate low-frequency trends in the data \cite{jebelli2018feasibility}. Following the filtering stages, baseline correction was performed by subtracting the minimum value of the signal to adjust for variations in the baseline level. To further prepare the data, the signal was normalized \cite{van2020standardized} to a scale between 0 and 1, which facilitated comparison across different datasets and ensured uniformity in subsequent analyses. 

Data analysis: The data analysis framework employs the cvxEDA algorithm \cite{choi2019feasibility} to decompose Electrodermal Activity (EDA) signals into their tonic and phasic components, enabling a detailed investigation of physiological responses. Since the phasic component represents rapid changes in the EDA signal, it is considered as the primary response to stimuli, reflecting the immediate physiological reactions triggered by external events or conditions. To ensure comparability across datasets, the phasic components of the signals were normalized. Since EDA signals were recorded at 4 Hz, they were aggregated by averaging every four consecutive post-warning data points to align with a 1-second interval, capturing participants' responses in 1-second increments. Stochastic inference: To compare the Electrodermal Response (EDR) between two activity conditions, Light and Moderate, two statistical tests were conducted to evaluate potential differences in the aggregated data. First, a paired t-test was performed on the combined datasets of each activity. This parametric test assesses whether there is a statistically significant difference in the means of the EDR responses between the two activities. In addition to the t-test, the Mann-Whitney U test, also known as the Wilcoxon rank-sum test, was conducted as a non-parametric alternative to compare the EDR responses between the Light and Moderate activity conditions. Unlike parametric tests, the Mann-Whitney U test evaluates whether the distributions of the two groups are significantly different by comparing their ranks rather than their means. 

To assess the importance of features derived from  (EDR) data across varying activity conditions, the EDR signals for "Light" and "Moderate" activities were processed to extract key statistical metrics. These metrics included the mean, standard deviation, variance, median, and peak amplitude of the EDR within the first 11 seconds post-warning. These features were selected to encapsulate critical physiological characteristics of the EDR signals and provide a comprehensive representation of participants' responses. Participant-level data for each metric were aggregated to compute group-level representations, ensuring a consistent basis for comparison. The aggregated data were then restructured into a feature matrix, with each metric serving as a feature. Feature importance was subsequently calculated by analyzing the normalized absolute values of the feature matrix for each activity condition to provide interpretable importance scores, indicating the relative contribution of each parameter.

\section{Result and Discussion}

Table \ref{table:mean_variance} presents a comparative analysis of the physical component of Electrodermal Response (EDR) over a 10-second post-warning interval for light and moderate activities. 
As shown in the table, the average EDR during light activity exhibits a steeper initial slope compared to moderate activity. This finding suggests that the perceived stress or sensitivity to warnings is higher during less physically demanding tasks, possibly due to lower baseline cognitive and physical engagement. Conversely, during moderate activity, the participants' physiological responses to warnings may be dampened, as the ongoing task might require greater attentional resources, leaving less capacity for additional stress responses. The average trends for light and moderate activities stabilize from the 5th to 6th seconds and 6th to 7th seconds of the interval, respectively. This stabilization indicates that the immediate impact of the warning dissipates over time, with the recovery period being dependent on activity intensity. This suggests that both the initial response to warnings and the recovery pattern differ based on activity intensity.

The paired t-test yielded a p-value of 0.0019, indicating a significant difference in the EDR between light and moderate activities ($p < 0.05$). This suggests that the intensity of the activity significantly impacts participants' stress levels following a warning. Additionally, the Mann-Whitney U test was conducted to validate these findings. The test produced a p-value of 0.0054, further confirming a significant difference between the two activities ($p < 0.05$).

\begin{table}[h]
\centering
\caption{Comparative analysis of post-warning EDR for light and moderate activities}
\label{table:mean_variance}
\begin{tabular}{cccc}
\toprule
\makecell{\textbf{Time} \\ \textit{(s)}} &
\makecell{\textbf{Light Activity} \\ \textit{Mean (Variance)}} &
\makecell{\textbf{Moderate Activity} \\ \textit{Mean (Variance)}} &
\makecell{\textbf{$\Delta$} }\\
\midrule
\makecell{[0--1]}   & 0.57 (0.03) & 0.61 (0.03) & 0.04 \\
\makecell{[1--2]}   & 0.65 (0.03) & 0.70 (0.03) & 0.05 \\
\makecell{[2--3]}   & 0.70 (0.03) & 0.76 (0.03) & 0.06 \\
\makecell{[3--4]}   & 0.73 (0.03) & 0.80 (0.03) & 0.07 \\
\makecell{[4--5]}   & 0.74 (0.03) & 0.83 (0.03) & 0.09 \\
\makecell{[5--6]}   & 0.75 (0.04) & 0.85 (0.03) & 0.10 \\
\makecell{[6--7]}   & 0.75 (0.06) & 0.86 (0.04) & 0.11 \\
\makecell{[7--8]}   & 0.75 (0.08) & 0.86 (0.04) & 0.11 \\
\makecell{[8--9]}   & 0.75 (0.09) & 0.86 (0.05) & 0.11 \\
\makecell{[9--10]}  & 0.75 (0.10) & 0.86 (0.06) & 0.11 \\
\bottomrule
\end{tabular}
\end{table}

\begin{figure}[!htb]
    \centering
    \includegraphics[width=0.40\textwidth]{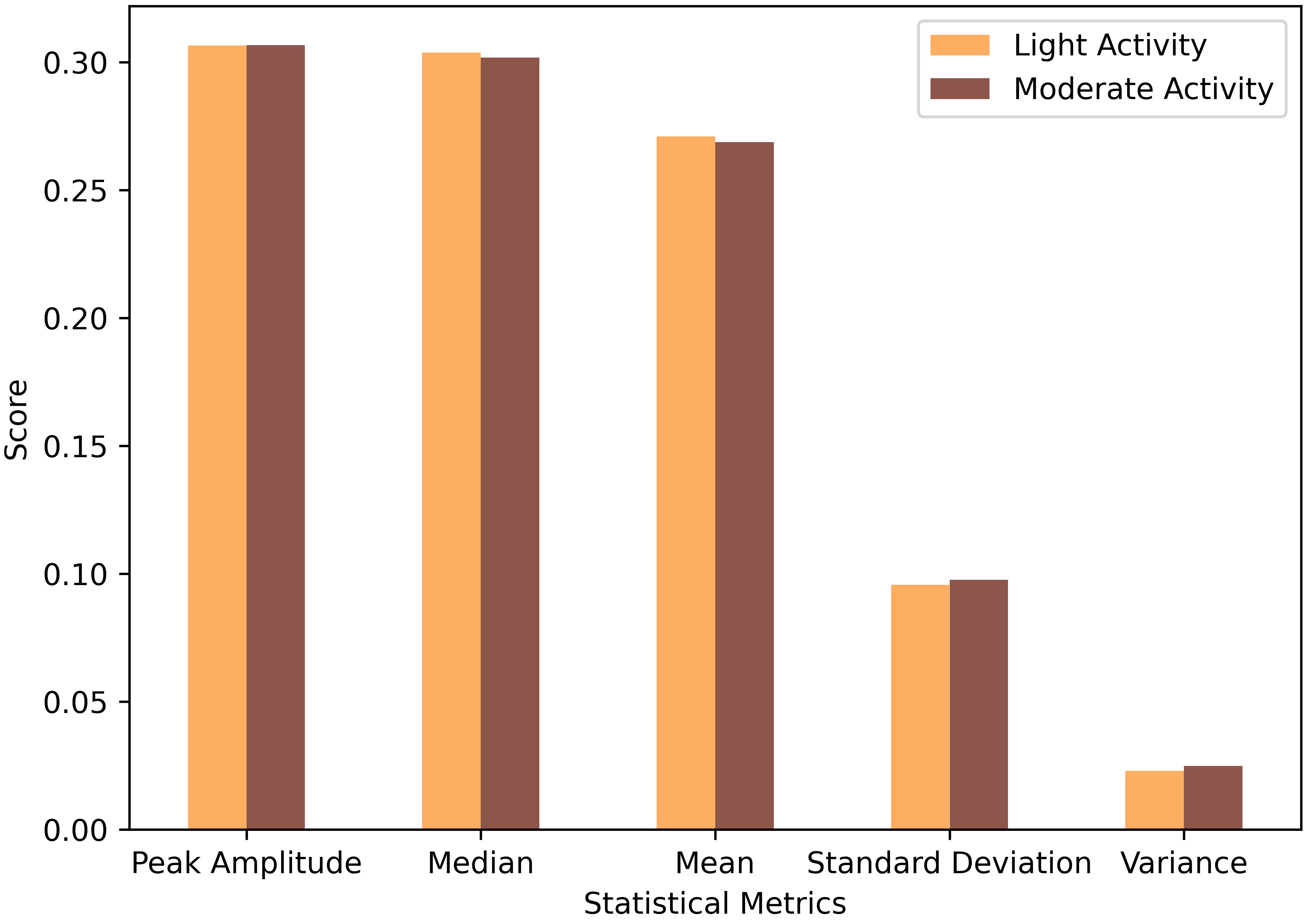}
    \caption{Feature importance for light and moderate activities}
    \label{fig_4}
\end{figure}

The feature importance analysis for statistical metrics derived from EDR data across light and moderate activities is summarized in Figure \ref{fig_4}. The results indicate that peak amplitude is the most critical feature in distinguishing physiological responses to warnings during both light and moderate activities. This suggests that peak amplitude, as a measure of the highest intensity of the Electrodermal Response (EDR), plays a key role in capturing the participants' stress or arousal levels post-warning. The median and mean metrics, which also exhibited high importance scores, provide additional insights into the central tendencies of EDR signals, highlighting their value in characterizing the overall physiological state of participants. Interestingly, the relatively low importance of standard deviation and variance suggests that variability in EDR signals is less effective compared to other statistic features of EDR signals. This could imply that the stress response captured by EDR is more consistently reflected in central tendency and peak metrics, rather than in the overall dispersion or variability of the data.

\section{Conclusion}
This study aimed to evaluate the stress responses of roadway workers exposed to warnings during routine roadway maintenance tasks, focusing on activities categorized as either light or moderate intensity. Physiological signals, specifically electrodermal activity (EDA), were recorded using a wristband sensor while participants performed highway maintenance tasks in an immersive virtual reality (VR) environment. This study introduces a human-sensing stress recognition model with two main contributions: (1) It leverages physiological data to assess workers' stress levels when receiving AR-enabled warnings that combine audio, visual, and haptic cues. (2) It enhances understanding of the interplay between stress and tasks involving physical exertion. These findings have important implications for enhancing work zone safety and the design of warning system. Tailoring warning systems to account for activity-specific stress responses could improve worker readiness and effectiveness in reacting to hazards. For instance, warnings during light activities may need to be less intense to avoid overstimulation, while warnings during moderate activities might need to be designed to capture workers' attention more effectively. The observed greater initial stress response among participants engaged in light activities highlights the interplay between activity intensity, stress levels, and risk perception following warnings. Furthermore, the importance of analyzing peak and central tendency metrics in EDR data is underscored, as they emerged as robust indicators of physiological responses across varying activity intensities. These insights pave the way for future research to explore the evolution of these metrics under diverse conditions and extend analyses to other physiological signals, thereby advancing our understanding of how to design more effective and inclusive warning systems.

This study identifies a few limitations that deserve attention in future research. Increasing the participant sample size and ensuring greater diversity would improve the generalizability of the results. Additionally, the study did not examine the potential influences of gender, race, age, or disability—factors that future studies could explore to gain deeper insight into stress recognition across diverse demographics. The use of head-mounted VR displays poses another limitation, as they may induce motion sickness or dizziness, potentially affecting stress measurements. Participant variability, such as differences in AR/VR familiarity and onsite work experience, could also impact outcomes, highlighting the need for subgroup analyses to account for these influences. Finally, this experiment focused solely on tasks of light and moderate intensity. Expanding future studies to include more physically demanding tasks could provide a more comprehensive understanding of how stress and activity levels affect stress responses.

\section {Acknowledgment}
The authors would like to acknowledge all participants in this study for their time and contributions. We also acknowledge the efforts of Sepehr Sabeti in developing the Virtual Reality model of the work zones and assisting with data collection, as well as Amit Kumar for his contributions to data collection.

\bibliographystyle{IEEEtran}
\bibliography{ASCE}  

\end{document}